\documentclass[12pt]{article}

\usepackage{amssymb}
\usepackage{amstext}

{

\def\w{\wedge}

\def\be{\begin{equation}}
\def\ee{\end{equation}}
\def\bea{\begin{eqnarray}}
\def\eea{\end{eqnarray}}

\begin{document}
\pagestyle{empty}
\begin{flushright}
\begin{tabular}{ll}
MCTP-05-85 & \\
ITFA-2005-28 & \\
hep-th/0507089& \\[0.3in]

\end{tabular}
\end{flushright}
\begin{center}
{\Large {\bf{Topological membrane theory from \\ \vspace{0.1cm} Mathai-Quillen formalism}}} \\ [.47in]
{{Lilia Anguelova$\, {}^{1}$, Paul de Medeiros$\, {}^{1}$ and Annamaria Sinkovics$\, {}^{2}$
}} \\ [.27in]
$\,{}^{1}$\, {\emph{Michigan Center for Theoretical Physics, Randall Laboratory, \\
University of Michigan, Ann Arbor, MI 48109-1120, U.S.A.}} \\ [.27in]
$\,{}^{2}$\, {\emph{Institute for Theoretical Physics, University of Amsterdam \\ Valckenierstraat 65, 1018 XE Amsterdam, The Netherlands.}} \\ [.27in]
{\tt{anguelov@umich.edu}}, {\tt{pfdm@umich.edu}}, {\tt{sinkovic@science.uva.nl}}
\\ [.45in]

{\large{\bf{Abstract}}} \\ [.2in]
\end{center}
It is suggested that topological membranes play a fundamental role in the
recently proposed topological M-theory.
We formulate a topological theory of membranes wrapping associative three-cycles in a seven-dimensional target space with $G_2$ holonomy.
The topological BRST rules and BRST invariant action are
constructed via the Mathai-Quillen formalism. In a certain gauge
we show this theory to be equivalent to a membrane theory
with two BRST charges found by Beasley and Witten. We argue that
at the quantum level an additional topological term should be
included in the action, which measures the contributions of
membrane instantons.
We construct a set of local and non-local observables for the topological membrane theory.
As the BRST cohomology of local operators turns out to be isomorphic to the de Rham cohomology of the $G_2$ manifold,
our observables agree with the spectrum of $d=4$, $N=1$ $G_2$ compactifications of M-theory.
\clearpage
\pagestyle{plain}
\pagenumbering{arabic}




\section{\large{Introduction}}

Topological string theory has been the source of many recent advances in mathematics and physics.
Among other things, it has led to new insights in BPS black hole entropy \cite{OSV} and in the structure of perturbative $N=4$ super Yang-Mills theory \cite{EW}.
It has also provided a possible quantum foam description of quantum gravity \cite{ORV}.

Its non-perturbative formulation, however, has only begun to be understood within the last few years.
An interesting development in this direction is the recently proposed topological M-theory \cite{M} (for other related  important work see also \cite{Z}.).
This theory is argued to provide a framework unifying the topological A- and B-models in a target space with one extra dimension, similarly to the relationship between physical M-theory and superstrings.
As is the case for physical M-theory, only the classical approximation of topological M-theory is presently well understood.
It is formulated in terms of the Hitchin action for seven-dimensional $G_2$ manifolds.
A similar construction for certain eight-dimensional $Spin(7)$ manifolds (which are a product of a Calabi-Yau threefold and a 2-torus) was also suggested in \cite{spin7}.
This eight-dimensional theory is inspired by the conjecture of a topological S-duality \cite{S} between the A- and B-models in six dimensions and so is proposed to be a topological analogy of F-theory, where the geometric structure of the extra two-torus encodes the S-duality transformations.

As in physical M-theory, one would expect the fundamental object of topological M-theory to be a membrane. In fact, in seven dimensions, a membrane is dual to a string. Thus an alternative microscopic description of topological M-theory may be provided by the topological $G_2$ string of \cite{BNS} (see also \cite{VSh} for an earlier investigation of the same theory).
This approach seems promising as it allows one to employ the powerful techniques of two-dimensional conformal field theory.
However, one is forced to define the theory, rather unconventionally, in terms of conformal blocks instead of the local operators of the untwisted sigma model.

In the present paper we explore the possibility for a fundamental formulation of topological M-theory
in terms of topological membranes.
Membranes wrapping associative three-cycles in a seven-manifold of $G_2$ holonomy contribute to the superpotential of the corresponding four-dimensional $N=1$ compactification of M-theory.
These contributions have been studied before in \cite{harmoo, beawit, AM}.
As is known from \cite{twistm}, the worldvolume theory of such membranes is automatically topologically twisted. We will describe below how these results fit in our formulation.

Our strategy is to consider the infinite-dimensional space of maps from a given three-manifold (the worldvolume of the membrane) to a fixed seven-dimensional target space of $G_2$ holonomy.
We then construct topologically invariant path integrals over such maps, which localize on the subspace of maps defining associative three-cycles in the $G_2$ manifold.
The formal properties of such path integrals will be derived using the Mathai-Quillen construction \cite{matqui, atijef}.

As will be reviewed in more detail in section 2, the basic ingredients of the Mathai-Quillen formalism are a vector bundle with a given section
and connection on the bundle specified.
This approach allows one to construct explicit representations of certain topological invariants of infinite-dimensional vector bundles (as path integrals localized on zeros of the chosen section).
Moreover, in the infinite-dimensional case, exterior derivatives acting on the bundle have the interpretation of nilpotent
{\emph{topological BRST}} operators, while the exponent in the Mathai-Quillen representation of the Euler class of the bundle can be viewed
as a BRST invariant action.
To illustrate this general framework, in section 3 we review two relevant examples of the Mathai-Quillen construction which describe
supersymmetric quantum mechanics (section 3.1) and the topological A-model (section 3.2) (see also \cite{swu, blatho}).

In section 4 we apply the Mathai-Quillen formalism to construct a topological theory of membranes.
The structure of the appropriate vector bundle is described in section 4.1 and the zeros of the chosen section are shown to correspond
to associative three-cycles.
The topological BRST rules and BRST invariant action for the theory are derived in section 4.2.
We find that a certain truncation of this theory obtained via a {\emph{static}} gauge choice is equivalent to a membrane theory
with two BRST charges found by Beasley and Witten \cite{beawit} (and also to the relevant \lq McLean multiplet' sector of the membrane theory
described by Harvey and Moore \cite{harmoo}).
In section 4.3 we find that the leading order terms in the bosonic part of the physical supermembrane action \cite{bersez},
expanded around an associative three-cycle, exactly match the leading order terms of the Mathai-Quillen action (in static gauge) only if one includes an additional topological term.
In section 4.4 we find that this topological term gives rise to precisely the membrane instanton factor proposed by Harvey and Moore \cite{harmoo} in the quantum theory.
We then construct local and non-local BRST invariant operators for the topological theory.
Finally, in section 5 we make some concluding remarks on remaining open issues and propose a membrane instanton expansion similar to the expansion for worldsheet instantons in topological string theory.

\vspace{0.5 cm}
\noindent
{\bf Note added:} While the typing of the present paper was being finalized, two preprints
\cite{memb1} appeared on the arXiv, which address the same idea of topological membranes providing the fundamental description of topological M-theory, but with approaches very different
from ours.


\section{\large{Mathai-Quillen formalism}} \label{sec-mq}
\setcounter{equation}{0}

We begin by reviewing the Mathai-Quillen construction {\cite{matqui}} of a set of representatives of certain de Rham cohomology classes on the base of a given vector bundle. See {\cite{blatho}} (and references therein) for an excellent review of this material in more detail. When the vector bundle in question is the tangent bundle of a finite-dimensional differentiable manifold, the Mathai-Quillen construction can be used to provide a family of different realizations of the Euler number of the manifold. The surprising result {\cite{atijef}} is that one can sometimes still use the Mathai-Quillen formalism to compute well-defined topological invariants when the given vector bundle is infinite-dimensional. For certain infinite-dimensional vector bundles, the Mathai-Quillen framework will allow us to relate a particular topological invariant called the {\emph{regularized Euler number}} to the partition function of an associated topological quantum field theory.


\subsection{Finite-dimensional case}

Consider an orientable vector bundle $E \rightarrow M$ over a compact orientable $m$-dimensional manifold $M$ (with coordinates $x^i$, $i=1,...,m$). It is assumed that the typical fibre of this bundle is a vector space of even dimension $2n \leq m$ (spanned by the orthonormal basis $f_a$, $a=1,...,2n$). $2n =m$ when $E$ is the tangent bundle $TM$ of an even-dimensional manifold.

The Euler class $e(E)$ of this bundle is an element of $H^{2n} (M,{\mathbb{R}})$ in de Rham cohomology.
For $E=TM$, the Euler number $\chi (E)$ is an integer obtained by simply evaluating $e(E) \in H^m (M,{\mathbb{R}})$ on $[M] \in H_m (M,{\mathbb{R}})$.
For $2n<m$ one obtains intersection numbers on $M$ by taking the wedge product of $e(E)$ with elements of $H^{m-2n} (M,{\mathbb{R}})$ before
evaluating on the fundamental cycle $[M]$.
Two seemingly very different representations of the Euler class that will be important in the forthcoming discussion are as follows.

The first representation is obtained as the Poincar\'{e} dual of the homology class of the discrete set of isolated zeros of a generic section ${\bar{s}}$ of $E$. For $E=TM$, evaluation of the Euler number just corresponds to counting (with signs) the number of zeros of the aforementioned generic vector field ${\bar{s}}$
\be
\chi (M) \; =\; \sum_{x, {\bar{s}}(x) =0} \pm 1 \; .
\label{euler1}
\ee
The expression above is the content of the Poincar\'{e}-Hopf theorem\footnote{A simple example of this result is for $M= S^2$. The well-known \lq hairy ball' theorem implies that every vector field on $S^2$ must vanish at two points. Both these points contribute with positive sign to give $\chi ( S^2 ) = 2$.}.
For a specific (i.e. non-generic) section $s$ of $TM$, the space of zeros $M_s$ of $s$ can have non-vanishing dimension. In this case one can show that the Euler number of $M$ is identical to the Euler number of the zero locus $M_s$. This property will be used to define a topological invariant for infinite-dimensional vector bundles in the next subsection.

The second representation can be computed if $E$ is equipped with a connection $\nabla = dx^i \nabla_i$ (defined by $\nabla_i f_a = \Gamma_{ia}^b f_b$ in terms of coefficients $\Gamma_{ia}^b$). The curvature $\Omega_{\nabla}$ of this connection can be understood as a two-form on $M$ valued in the space of $2n$$\times$$2n$ antisymmetric matrices. A representative $e_\nabla (E)$ of the Euler class $e(E)$ is proportional to the matrix Pfaffian of $\Omega_\nabla$
\footnote{Recall that the Pfaffian of a $2n$$\times$$2n$ antisymmetric matrix $A^{ab}$ is defined to be the number
${\mbox{Pf}} (A) = ( (-1)^n / 2^n n! ) \epsilon_{a_1 ... a_{2n}} A^{a_1 a_2} ... A^{a_{2n-1} a_{2n}}$ and obeys
$( {\mbox{Pf}} (A) )^2 = {\mbox{det}} (A)$. Thus ${\mbox{Pf}} ( \Omega_\nabla )$ is a closed $2n$-form on $M$ as required (it is closed as a result
of the Bianchi identity $\nabla_{[i} \Omega_{\nabla\, jk]}^{ab} = 0$ for $\Omega_\nabla$).}.
This can be used to show that the cohomology class of $e_\nabla (E)$ is independent of the choice of connection $\nabla$. Using the well-known representation of the Pfaffian in terms of Berezin integrals allows us to write
\be
e_\nabla (E) \; =\; (2\pi )^{-n} \, {\mbox{Pf}}( \Omega_\nabla ) \; =\; (2\pi )^{-n} \int d \chi \, e^{\frac{1}{2} \Omega_\nabla^{ab} \chi_a \chi_b}  \; ,
\ee
where the second equality involves $2n$ Grassmann odd variables $\chi_a$ and Berezin measure $d \chi = d \chi_1 ... d \chi_{2n}$ , defined such that $\int d \chi_a \, \chi_b = \delta_{ab}$. For $E=TM$, evaluation of the Euler number corresponds to integrating $e_\nabla (E)$ over $M$
\be
\chi (M) \; =\; \int_M e_\nabla (E) \; .
\label{euler2}
\ee
This is the Gauss-Bonnet theorem.

The Mathai-Quillen approach allows one to generalise the formulae above to construct an explicit representative $e_{\nabla , s} (E)$ of the same Euler class $e(E)$ in terms of both a connection $\nabla$ and a section $s$ on $E$. Given this data, the representative is
\be
e_{\nabla ,s} (E) \; =\; (2\pi )^{-n} \int d \chi \, e^{-\frac{1}{2} g_{ab} s^a s^b  +i\, dx^i ( \nabla_i s^a ) \chi_a + \frac{1}{2} \Omega_\nabla^{ab} \chi_a \chi_b}  \; ,
\label{mq1}
\ee
where $g_{ab}$ is the (flat) fibre metric and $\nabla_i s^a = \partial_i s^a + \Gamma_{ib}^a s^b$. Closure of this $2n$-form on $M$ can be shown without explicitly performing the Berezin integrals provided one defines
\footnote{Exterior derivation on $\chi_a$ here should not be confused with the Berezin measure $d \chi$ defined previously.}
\be
d \chi_a \; =\; i\, g_{ab} s^b \; .
\ee
It is assumed $d s^a = dx^i \nabla_i s^a$ and that $dx^i$ anticommutes with $\chi_a$. It is therefore convenient to define the closed, Grassmann odd variable $\psi^i = dx^i$. When $E=TM$ one obtains the same Euler number $\chi (M)$ by integrating $e_{\nabla ,s} (E)$ over $M$ for any choice of $\nabla$ and $s$. Having replaced $dx^i$ by $\psi^i$, the integration of a top form $\int_M \omega_{(m)}$ (where $\omega_{(m)} = 1/m!\, \omega_{i_1 ... i_m} (x) dx^{i_1} \wedge ... \wedge dx^{i_m}$) mentioned above should be replaced by $\int_M dx \int d \psi \, {\cal{O}}_{\omega_{(m)}}$, where $\int_M dx$ is Lebesgue integration over $M$, $d \psi = d \psi^{i_1} ... d \psi^{i_m}$ is the Berezin measure defined by $\int d \psi^i \, \psi^j = \delta^{ij}$ and ${\cal{O}}_{\omega_{(m)}} = 1/m!\, \omega_{i_1 ... i_m} (x) \psi^{i_1} ... \psi^{i_m}$ is a function on $M$ obtained from $\omega_{(m)}$ by the replacement.

Let us conclude the finite-dimensional discussion by showing how the Mathai-Quillen representative $e_{\nabla ,s} (E)$ provides a beautiful interpolation between the two expressions for the Euler number of $E=TM$ described above. Since the Euler number is the same for any choice of section, consider the one parameter family of representatives $e_{\nabla , \gamma {\bar{s}}} (E)$ obtained by replacing $s \rightarrow \gamma {\bar{s}}$ in ({\ref{mq1}}) for all $\gamma \in {\mathbb{R}}$ and generic section ${\bar{s}}$. Evidently, we recover the analytic Gauss-Bonnet formula ({\ref{euler2}}) in the $\gamma \rightarrow 0$ limit since $e_{\nabla ,s=0} = e_{\nabla}$. In the opposite $\gamma \rightarrow \infty$ limit it is clear that the $-1/2 \, \gamma^2 ||{\bar{s}}||^2$ term dominates the exponent in ({\ref{mq1}}) and that $e_{\nabla , \gamma {\bar{s}}} (E)$ can only be non-vanishing at points where ${\bar{s}} =0$. We should therefore only expect contributions to the Euler number integral from the discrete zero locus of ${\bar{s}}$. The naive analysis above can be made precise using a stationary phase approximation (which turns out to be exact in this simple case) to show that, in the $\gamma \rightarrow \infty$ limit, the Euler number is indeed computed by just adding up (with signs) the zeros of ${\bar{s}}$. Thus we have recovered the topological Poincar\'{e}-Hopf result ({\ref{euler1}}).


\subsection{Infinite-dimensional case}

Although a very elegant way to relate classical expressions for the Euler number, thus far the Mathai-Quillen formalism has not led to any new topological information about differentiable manifolds. This is due to the fact that all $e_{\nabla ,s} (E)$ are representatives of the same cohomology class $e(E) \in H^{2n} (M,{\mathbb{R}})$ when $E$ is finite-dimensional. For the case of infinite-dimensional vector bundles, many of the classical expressions found above are no longer defined (formally we would be dealing with \lq infinite forms'). Indeed $e_\nabla (E)$ has no meaning in this case. The trick is to recall the localization that occured in the Poincar\'{e}-Hopf formula in the finite-dimensional tangent bundle case. As explained in {\cite{atijef}}, for the case of infinite-dimensional vector bundles one can obtain new and well-defined topological information about the bundle $E \rightarrow M$ via the Mathai-Quillen construction provided one chooses an appropriate section $s$ whose zero locus $M_s$ is a finite-dimensional submanifold of $M$. Given such a setup, the idea is to {\emph{define}} the {\emph{regularized Euler number}} $\chi_s (M)$ of the infinite-dimensional tangent bundle $E=TM$ to be
\be
\chi_s (M) \; =\; \chi ( M_s ) \; .
\ee
The Euler number on the right hand side being well-defined for $M_s$ compact and finite-dimensional. Unlike the finite-dimensional case, this definition clearly depends on the choice of non-generic section $s$. As will be seen in the forthcoming examples, there is typically a section which is naturally associated with most choices of infinite-dimensional vector bundle.

The regularized Euler number has a formal Mathai-Quillen type expression in terms of the \lq partition function' path integral
\be
\chi_s (M) \; =\; ( 2\pi )^{-{\scriptsize{\mbox{dim}}}( M_s ) /2}\int_M [dx] \, [d \psi ] \, [ d \chi ] \, e^{- I_{\nabla , s}}  \; ,
\label{mq2}
\ee
where the exponent is now an \lq action' functional
\be
I_{\nabla ,s} \; =\; \int \frac{1}{2} \, g_{ab} s^a s^b  -i\, \psi^i ( \nabla_i s^a ) \chi_a - \frac{1}{4} \Omega_{\nabla\, ij}^{ab} \psi^i \psi^j \chi_a \chi_b \; .
\label{act}
\ee
Note that the formal index contractions will typically involve integrals since they are summed over an infinite number of dimensions. The exterior derivative on this infinite-dimensional space can be interpreted as a topological BRST operator $\delta$. Indeed one can check that the action $I_{\nabla ,s}$ is invariant under the transformations
\be
\delta s^a \; =\; \psi^i \nabla_i s^a \; , \quad\quad \delta \psi^i \; =\; 0  \; , \quad\quad \delta \chi_a \; =\; i\, g_{ab} s^b \; .
{\label{brst}}
\ee
These transformations are nilpotent up to equations of motion derived from $I_{\nabla ,s}$ and rotations in the tangent space $T_x M$. This defines a so called equivariant cohomology associated with $\delta$ though this subtlety will not be important in the forthcoming analysis. We will now discuss some concrete examples which realize the aforementioned structure.


\section{\large{Examples}} \label{sec-Amodel}
\setcounter{equation}{0}

We will now highlight how the Mathai-Quillen construction applies to some well-studied topological theories concerned with counting certain maps
from low-dimensional compact manifolds to higher dimensional ones. Both of the manifolds are finite-dimensional though the space of maps between
them has infinite dimension. The examples in question are supersymmetric quantum mechanics and the topological A-model.
Most of this material is covered in more detail in {\cite{blatho,swu}}. These subjects are very interesting in their own right and there is a huge
literature on both (see for example {\cite{blatho,wit1,wit2}} and references therein).
We include this brief summary only to facilitate readers, who need to gain familiarity with the formalism, before turning to the construction of topological membranes in section 4.


\subsection{Supersymmetric quantum mechanics}

In this case the infinite-dimensional vector bundle is the tangent bundle of the space of loops in a finite-dimensional compact Riemann manifold $X$.
That is, $E=TM$ and $M = {\mbox{Map}} ( S^1 , X)$.
Each coordinate $x= \{ x^i (t) \}$ in $M$ corresponds to a set of coordinates parameterizing a loop in $X$ (the parameter in question being the coordinate
$t \in [0,1]$ on $S^1$, and hence $x^i (1) = x^i (0)$).
Similarly a tangent vector $v = \{ v^i (x(t)) \}$ at a given point $x \in M$ corresponds to a set of tangent vectors based along points of the loop in $X$.
The metric $g_{ij}$ on $X$ induces a metric on $M$ defined such that at each point $x \in M$
\be
g_x ( u , v ) \; =\; \int_0^1 dt\, g_{ij} (x(t)) u^i (x(t)) v^j (x(t)) \; ,
\ee
for any two tangent vectors $u, v \in T_x M$. In this way a general differential form on $X$ induces a differential form on $M$.

Perhaps the simplest section of the bundle $TM$ corresponds to the vector field on $M$ generating constant shifts
$x^i (t) \rightarrow x^i (t+\epsilon )$ around the loops. This field is made up of the vectors ${\dot{x}} = \{ {\dot{x}}^i (t) \} \in T_x M$ at each
$x \in M$. With this choice of section, the Mathai-Quillen formalism implies that the generalized Euler number of $M$ is simply the Euler number of $X$
\be
\chi_{\dot{x}} (M) \; =\; \chi (X) \; .
\ee
The reason being that the zero locus $M_{\dot{x}}$ of ${\dot{x}}$ corresponds to the space of constant maps into $X$ which is just isomorphic to $X$ itself.

That the expression above also follows from the Mathai-Quillen path integral described previously can be rigorously proven via powerful localization
theorems which apply for this simple theory (see e.g. {\cite{blatho}}). We will now describe the explicit form of the action and BRST transformations
for this theory which follow from the Mathai-Quillen formalism.

By introducing the vielbein $e_i^a$ associated with the metric $g_{ij}$, one can write
the chosen section $s^a = e_i^a \, {\dot{x}}^i$ and the variable $\chi_a = e_a^i \, {\bar{\psi}}_i$ in terms of another Grassmann odd variable ${\bar{\psi}}_i$
(for each value of $t$).
It is worth making this distinction as an orthonormal basis was used for vectors in the previous section.
In terms of these variables, the action ({\ref{act}}) is given by
\be
I_{SQM} \; =\; \int_0^1 dt \, \left( \frac{1}{2} \, g_{ij} \, {\dot{x}}^i {\dot{x}}^j +i\, {\bar{\psi}}_i  \nabla_t \psi^i  - \frac{1}{4} R_{ijkl} \psi^i \psi^j {\bar{\psi}}^k {\bar{\psi}}^l \right) \; ,
\ee
where $\nabla_t \psi^i = {\dot{\psi}}^i + \Gamma^i_{jk} \psi^j {\dot{x}}^k$ is defined by the action of the Levi-Civita connection $\nabla$ of $g$ pulled
back to the loop via the map $x$ and $R$ is the Riemann tensor of $g$.

The action above is invariant under the topological BRST transformations
\be
\delta x^i \; =\; \psi^i \; , \quad\quad \delta \psi^i \; =\; 0  \; , \quad\quad \delta {\bar{\psi}}_i \; =\;  i\, g_{ij} {\dot{x}}^j + \Gamma^k_{ij} \psi^j {\bar{\psi}}_k \; ,
\ee
which just follow from ({\ref{brst}})
\footnote{The extra $\Gamma^k_{ij} \psi^j {\bar{\psi}}_k$ term appearing in $\delta {\bar{\psi}}_i$ follows from the change of basis
$\chi_a = e_a^i \, {\bar{\psi}}_i$ employed relative to ({\ref{brst}}). In particular, the BRST transformation
$\delta e_a^i = - \Gamma_{jk}^i \psi^j e^k_a$ of the inverse vielbein has been used
(which follows from $\delta x^i = \psi^i$ using $\delta_{ab} = e_a^i e_b^j g_{ij}$).}
\footnote{Notice that $\psi^i$ is a function of $t$ but not $x^i(t)$. In particular $\partial_i \psi^j =0$. This means that $\delta x^i = \psi^i$
does not correspond to a general coordinate transformation on $X$. Indeed the components $T(x)$ of an arbitrary tensor on $X$ have the BRST transformation
$\delta T(x) = \psi^i \partial_i T(x)$.}.
These transformations satisfy $\delta^2 =0$ on-shell
\footnote{The use of fermionic equations of motion is required to close the BRST algebra. There is a very straightforward way to close the algebra
off-shell using auxiliary fields though we will not discuss this here.}.

The action $I_{SQM}$ is a trivial element in the cohomology of $\delta$ because
\be
I_{SQM} \; =\; \delta \Psi_{SQM} \; =\; - \frac{i}{2} \, \delta \left( \int_0^1 dt \, {\bar{\psi}}_i \, {\dot{x}}^i \right) \; ,
\label{exact}
\ee
on-shell.
If one were to include a coupling constant multiplying $I_{SQM}$ then formal arguments imply the corresponding Mathai-Quillen path
integral (with BRST invariant operator insertions) will be independent of the value of this constant due to ({\ref{exact}}).
It is therefore typically convenient to do computations in the limit where the constant is large so that only the minima of the action
$I_{SQM}$ contribute to the path integral.
Notice that the critical points of $\delta$ minimize $I_{SQM}$ and precisely correspond to the constant maps on which the
path integral localizes. This is a generic feature of the Mathai-Quillen construction.

It is worth noting that $I_{SQM}$ has an additional ${\mathbb{Z}}_2$ symmetry generated by exchanging $\psi^i \leftrightarrow {\bar{\psi}}^i$ with
$x^i$ left invariant.
One can show that all the cohomological properties above are unaffected by making this change of variables in the BRST rules
(together with the relabeling $\delta \rightarrow {\bar{\delta}}$).
In particular, $I_{SQM}$ is closed and exact under the resulting nilpotent transformations ${\bar{\delta}}$,
\be
{\bar{\delta}} x^i \; =\; {\bar{\psi}}^i \; , \quad\quad {\bar{\delta}} {\bar{\psi}}^i \; =\; 0  \; , \quad\quad {\bar{\delta}} \psi_i \; =\; i g_{ij} {\dot{x}}^j + \Gamma_{ij}^k {\bar{\psi}}^j \psi_k \; .
\ee
The theory above is therefore just $N=2$ supersymmetric quantum mechanics on $X$.
A similar discrete symmetry will be found to exist when we come to consider a topological theory of membranes.


\subsection{Topological sigma models : the A-model}

In the second example we consider, $M$ is the space ${\mbox{Map}} ( \Sigma_2 , X)$ of maps $\phi$ from a Riemann surface $\Sigma_2$ to a Calabi-Yau manifold $X = {\mbox{CY}}_3$
\footnote{Strictly speaking, at this stage we only require $X$ to be a symplectic manifold with almost complex structure compatible with the complex structure of $\Sigma_2$.
The choice of $X = {\mbox{CY}}_3$ will guarantee that certain anomalies vanish in the corresponding topological quantum field theory.}.
The topological model we will describe can be understood as a certain {\emph{twisted}} version of an $N=(2,2)$ superconformal non-linear sigma model with target space $X = {\mbox{CY}}_3$.
There are two inequivalent twists which give rise to topological theories called the {\emph{A-}} and {\emph{B-models}}.
The A-model can also be understood via the Mathai-Quillen formalism for a certain choice of section; the details of this construction will be summarized shortly.
See {\cite{swu}} for a more thorough analysis of the Mathai-Quillen construction for the A-model.
Whether there is a Mathai-Quillen formulation of the B-model too, is not quite clear at present and we will not delve into this topic, although we mention briefly below the main source of difficulty.

If we introduce complex coordinates $(z,{\bar{z}}) \in \Sigma_2$ , then each coordinate
$\phi = \{ \phi^I (z,{\bar{z}}) \} = \{ \phi^i (z,{\bar{z}}) , \phi^{\bar i} (z,{\bar{z}}) \} \in M$ corresponds to a Riemann surface embedded in $X$
\footnote{$I$ is a real 6-vector index on $X$ which it will sometimes be convenient to split into 3 (anti-)holomorphic indices $({\bar{i}})i$.}.
The metric tensor and differential forms on $X$ induce their counterparts on $M$ via a straightforward generalization of the method described in the previous subsection.

Let us now concentrate on the A-model. The infinite-dimensional space of maps $M = {\mbox{Map}} ( \Sigma_2 , X)$ is localized on the finite-dimensional space $M_s$ of maps
$\{ \phi^i (z) , \phi^{\bar i} ({\bar{z}}) \}$ corresponding to Riemann surfaces which are holomorphically embedded in $X$.
Of course, this space of {\emph{holomorphic maps}} is spanned by solutions of the equations
\be
\partial_{\bar{z}} \phi^i \; =\; 0 \; , \quad\quad \partial_{z} \phi^{\bar{i}} \; =\; 0 \; ,
\ee
which can be understood as the zeros of a map $s : ( \phi^i , \phi^{\bar{i}} ) \rightarrow ( \partial_{\bar{z}} \phi^i , \partial_{z} \phi^{\bar{i}} )$.
The map $s$ is a section of the bundle $E$ whose typical fibre at $\phi$ is the vector space $\Gamma ( T^{*(0,1)} \Sigma_2 \otimes \phi^* T^{(1,0)} X)$, i.e. the space
of sections of the bundle $T^{*(0,1)} \Sigma_2 \otimes \phi^* T^{(1,0)} X$.
Thus $E$ is not the tangent bundle of $M$ in this case but, as required, its fibre dimension is less than the dimension of $M$ {\cite{swu}}.
The difficulty in realizing the B-model a la Mathai-Quillen is that the naive formulation leads to a bundle whose fibre dimension exceeds the
dimension of $M$.

The Mathai-Quillen section $s^a$ is therefore identified with $( \partial_{\bar{z}} \phi^i , \partial_{z} \phi^{\bar{i}} )$ whilst the variables
$\psi^i$ and $\chi_a$ are written in terms of the fermionic fields $\psi^I$ and $( \chi_{\bar{z}}^i , \chi_{z}^{\bar{i}} )$ on $\Sigma_2$
\footnote{Appropriate vielbeins are implicitly included in this identification.}.
In terms of these variables, the Mathai-Quillen action is given by
\be
I_A \; =\; \int_{\Sigma_2} d^{\,2} z \, \left( g_{i{\bar i}} \, \partial_{\bar{z}} \phi^i \partial_{z} \phi^{\bar{i}} +i\, g_{i{\bar i}} \left( \chi_{\bar{z}}^i  \nabla_z \psi^{\bar i} + \chi_{z}^{\bar i}  \nabla_{\bar z} \psi^i  \right) - R_{i{\bar i}j{\bar j}} \, \chi_{\bar{z}}^i \chi_{z}^{\bar i} \psi^j \psi^{\bar j} \right) \; ,
\label{actiona}
\ee
where $\nabla_z \psi^{\bar i} = \partial_z \psi^{\bar i} + \Gamma^{\bar i}_{{\bar j}{\bar k}} \psi^{\bar j} \partial_z \phi^{\bar k}$ and its complex conjugate
$\nabla_{\bar z} \psi^i$ are defined by the action of the Levi-Civita connection on $X$ pulled back to $\Sigma_2$ by $\phi$.
$I_A$ has a global $U(1)$ {\emph{ghost number}} symmetry at the classical level under which $( \phi , \psi , \chi )$ have charges $(0,1,-1)$.

The action ({\ref{actiona}}) is invariant under the BRST transformations
\be
\delta \phi^I \; =\; i \psi^I \; , \quad\quad \delta \psi^I \; =\; 0  \; , \quad\quad \delta \chi_z^{\bar i} \; =\; - \partial_z \phi^{\bar i} -i \Gamma^{\bar i}_{{\bar j}{\bar k}} \psi^{\bar j} \chi_z^{\bar k} \; ,
\label{brsta}
\ee
which again follow from ({\ref{brst}}) and are nilpotent up to fermionic equations of motion.
It is worth noting that the A-model one obtains by twisting the superconformal sigma model {\cite{wit1}} has two nilpotent scalar supersymmetries.
The transformations ({\ref{brsta}}) follow from the twisted model
after setting the two scalar supersymmetry parameters equal (this is done in {\cite{wit2}} to simplify the analysis of the BRST cohomology).

The action $I_A$ is also $\delta$-exact since
\be
I_A \; =\; \delta \Psi_A \; =\; \delta \left( -\frac{1}{2} \int_{\Sigma_2} d^{\, 2} z \, g_{i{\bar i}} \left( \chi_{\bar z}^i \partial_z \phi^{\bar i} + \chi_{z}^{\bar i} \partial_{\bar z} \phi^{i}  \right) \right) \; ,
\label{exacta}
\ee
on-shell. Again if one introduces a coupling constant $2t$ multiplying $I_A$ then ({\ref{exacta}}) formally implies that the Mathai-Quillen path
integral does not depend on $t$ and only gets contributions from minima of $I_A$ in the large $t$ limit.
Not surprisingly, the critical points of $\delta$ precisely correspond to holomorphic maps which minimize $I_A$.

There is a topological term one can add to the Mathai-Quillen action given by the integral over $\Sigma_2$ of the pull back via $\phi$ of the
K\"{a}hler form $K$ on $X$. This positive definite term
\be
\int_{\Sigma_2} \phi^* (K) \; =\; \int_{\Sigma_2} d^{\, 2} z\, g_{i{\bar i}} \left( \partial_z \phi^i \partial_{\bar z} \phi^{\bar i} - \partial_{\bar z} \phi^i \partial_z \phi^{\bar i}  \right) \; ,
\label{top1}
\ee
is invariant under diffeomorphisms of $X$ and is therefore also BRST invariant. In fact ({\ref{top1}}) depends only on the homotopy class of $\phi$ and
the cohomology class of $K$. Indeed, it is quantized and equal to $2\pi k$ (for some non-negative integer $k$) if $H^2 (X,{\mathbb{Z}}) = {\mathbb{Z}}$.
It is natural to include such a term from the perspective of the twisted sigma model since the purely bosonic part of the sigma model
action (which is not affected by the twist) can be written as
\be
\frac{1}{2} \int_{\Sigma_2} d^{\, 2} z \, g_{IJ} \, \partial_z \phi^I \partial_{\bar z} \phi^J \; =\; \int_{\Sigma_2} d^{\,2} z \, g_{i{\bar i}} \, \partial_{\bar{z}} \phi^i \partial_{z} \phi^{\bar{i}} + \frac{1}{2} \int_{\Sigma_2} \phi^* (K) \; .
\ee

Although adding ({\ref{top1}}) does not modify the equations of motion of the classical theory, its inclusion at the quantum level is found to be crucial in
the analysis of the twisted A-model {\cite{wit1}}. To illustrate how this works, consider the expectation value (or \lq observable')
\be
\langle {\cal{O}}_{p_1} ... {\cal{O}}_{p_s} \rangle (t) \; =\; \int_M [d \phi ] \, [d \psi ] \, [d \chi ] \, {\cal{O}}_{p_1} ... {\cal{O}}_{p_s} \, e^{-t S_A} \; ,
\label{obs1}
\ee
of a set of BRST invariant local operators $\{ {\cal{O}}_{p_r} | r=1,...,s \}$ where ${\cal{O}}_{p_r}$ has ghost number $p_r$ and is evaluated at
a point $( z_r , {\bar z}_r ) \in \Sigma_2$ such that $\phi ( z_r , {\bar z}_r )$ lies on a homology $p_r$-cycle in $X$.
The path integral above is with respect to the full action
\be
t\, S_A \; =\; 2t\, I_A + t \int_{\Sigma_2} \phi^* (K) \; .
\ee
For a fixed Calabi-Yau manifold $X$ with K\"{a}hler form $K$, the topological term $\int_{\Sigma_2} \phi^* (K)$ in ({\ref{obs1}}) can only be factored
out if one restricts the path integral to be over that portion of configuration space $M_k \subset M$ corresponding to maps $\phi$ with the same
homotopy class (which must also obey the constraints $\phi ( z_r , {\bar z}_r ) \in H_{p_r} (X,{\mathbb{R}})$ here).
Let us assume we have a Calabi-Yau geometry such that the quantization $\int_{\Sigma_2} \phi^* (K) = 2\pi k$ occurs so that these classes
are labeled by the non-negative integer $k$.
Formally one can therefore write $\langle {\cal{O}}_{p_1} ... {\cal{O}}_{p_s} \rangle (t)$ as a sum over $k$ of the path integrals
\be
\langle {\cal{O}}_{p_1} ... {\cal{O}}_{p_s} \rangle_k \; =\; \int_{M_k} [d \phi ] \, [d \psi ] \, [d \chi ] \, {\cal{O}}_{p_1} ... {\cal{O}}_{p_s} \, e^{-2t I_A} \; ,
\ee
each weighted by a {\emph{worldsheet instanton}} factor $e^{-2\pi k t}$.
Each path integral $\langle {\cal{O}}_{p_1} ... {\cal{O}}_{p_s} \rangle_k$ is of the Mathai-Quillen form (although over the reduced configuration space $M_k \subset M$) and is independent of $t$.
It turns out that the entire $t$ dependence of $\langle {\cal{O}}_{p_1} ... {\cal{O}}_{p_s} \rangle (t)$ comes from the instanton factors in this way.

It can be shown that the measure of this restricted path integral is not invariant under the classical ghost number symmetry of the A-model.
Indeed one can compute the anomaly to be $3 \chi ( \Sigma_2 )$ which is the dimension of the space of solutions of
$\nabla_{\bar z} \psi^i = \nabla_z \psi^{\bar i} = 0$ via the Riemann-Roch theorem (and is independent of $k$).
One finds that the observables $\langle {\cal{O}}_{p_1} ... {\cal{O}}_{p_s} \rangle_k$ are computable as discrete sums precisely when the ghost number of
${\cal{O}}_{p_1} ... {\cal{O}}_{p_s}$ equals the aforementioned anomaly so that the overall ghost number is zero, i.e.
\be
3 \chi ( \Sigma_2 ) \; =\; \sum_{r=1}^s p_r \; .
\label{anom}
\ee
Indeed, this implies that the reduced configuration space $M_k$ (supplemented with the $\phi ( z_r , {\bar z}_r ) \in H_{p_r} (X,{\mathbb{R}})$ constraints) has
dimension zero and the path integral $\langle {\cal{O}}_{p_1} ... {\cal{O}}_{p_s} \rangle_k$ simply counts the number $| M_k |$ of points in this space.
Typically the difference between the left and right hand sides of ({\ref{anom}}) gives the dimension of the space $M_k$ over which one integrates
the localized path integral.

Thus generically one finds
\be
\langle {\cal{O}}_{p_1} ... {\cal{O}}_{p_s} \rangle (t) \; =\; \sum_{k=0}^{\infty} e^{-2\pi k t} \, | M_k | \; ,
\ee
when the ghost number anomaly equation above is satisfied.
Coupling the A-model to topological gravity on $\Sigma_2$ implies that it has the structure of a topological theory of closed strings.
The observables of this topological string theory correspond to the Gromov-Witten invariants.

With regard to some of the points made above it may now be helpful to make some side remarks concerning the general structure of observables.
It is assumed that $\langle \delta \Upsilon \rangle =0$ for any operator $\Upsilon$, since (using $\delta S_A = 0$) the variation $\delta \left( \Upsilon \, e^{-t S_A} \right)$
just corresponds to a \lq total derivative' in configuration space.
Thus any two BRST invariant operators which differ by $\delta \Upsilon$ give the same observable.
Hence the classification of observables corresponds to computing the cohomology of $\delta$.
Recall that $\delta$ can be understood as the exterior derivative on $M$ but is also related to the exterior derivative $d$ on $X$ via the simple formula
\be
\delta {\cal{O}}_{\omega_p} \; =\; i\, {\cal{O}}_{d \omega_p } \; ,
\ee
where the local operator ${\cal{O}}_{\omega_p} = 1/p! \, \omega_{I_1 ... I_p} (\phi ) \psi^{I_1} ... \psi^{I_p}$ at the point $(z, {\bar z}) \in \Sigma_2$ (and thus $\phi (z, {\bar z}) \in X$) is related to the
$p$-form $\omega_p = 1/p! \, \omega_{I_1 ... I_p} (\phi ) d\phi^{I_1} \wedge ... \wedge d\phi^{I_p}$ by just replacing $d\phi^I$ with $\psi^I$.
${\cal{O}}_{\omega_p}$ is therefore BRST invariant if $\omega_p$ is closed.
A natural set of closed $p$-forms on $X$ consists of elements $\Gamma_p$ of $H^p (X,{\mathbb{R}})$ that are Poincar\'{e} dual to
(and have delta function support on) homology $p$-cycles in $H_p (X,{\mathbb{R}})$.
The corresponding BRST invariant local operators ${\cal{O}}_{\Gamma_p}$ have ghost number $p$ and are at $(z, {\bar z}) \in \Sigma_2$ thus having
$\phi (z, {\bar z})$ at a point on the dual homology $p$-cycles (as was assumed above).
This provides an isomorphism between BRST cohomology of local operators in the A-model and de Rham cohomology on $X$.

Non-local operators defined on homology cycles of $\Sigma_2$ can be constructed from a local operator ${\cal{O}}^{(0)}$ via so called {\emph{descent equations}}.
The topological structure of the theory implies that expectation values of local operators should be independent of the point on $\Sigma_2$ they are evaluated at.
This implies the first descent equation $d {\cal{O}}^{(0)} = \delta \xi_1$ for some $1$-form $\xi_1$ on $X$.
The corresponding BRST invariant operator is ${\cal{O}}^{(1)} = \int_{\gamma_1} \phi^* ( \xi_1 )$ for any one-cycle $\gamma_1 \in H_1 ( \Sigma_2 , {\mathbb{R}})$.
One can continue in this way to construct a final BRST invariant operator ${\cal{O}}^{(2)} = \int_{\Sigma_2} \phi^* ( \xi_2 )$ from the descent equation
$d \xi_1 = \delta \xi_2$ for some $2$-form $\xi_2$ on $X$ (for example, one such $\xi_2 = K$ wherein ${\cal{O}}^{(2)}$ is the topological term we added to the action $I_A$).


\section{\large{Topological membranes}} \label{sec-tm}
\setcounter{equation}{0}

We will now follow the Mathai-Quillen construction to formulate a topological theory with $M = {\mbox{Map}} ( \Sigma_3 , X)$
being the space of maps $x$ from a three-manifold $\Sigma_3$ to a seven-dimensional manifold $X$ of $G_2$ holonomy
(defined by the existence of a closed and coclosed associative 3-form $\Phi$).
Introducing coordinates $\sigma^a \in \Sigma_3$, where $a=1,2,3$, then each coordinate
$x = \{ x^I ( \sigma ) \} \in M$, where $I=1,...,7$, corresponds to a Riemannian three-manifold or \lq membrane worldvolume' embedded in $X$.

For each such embedded membrane $x( \Sigma_3 )$ there exists a natural decomposition of the tangent bundle $TX = Tx( \Sigma_3 ) \oplus Nx( \Sigma_3 )$
in terms of the tangent and normal bundles of $\Sigma_3$ in $X$.
This decomposition is orthogonal with respect to the metric induced on $\Sigma_3$ from $X$.
At some instances in the forthcoming discussion it will be convenient to implement this splitting at each point in $M$ such that coordinates
$x^I ( \sigma )$ in $X$ are written in terms of coordinates $( x^a (\sigma ) , x^i (\sigma ) )$, where $a=1,2,3$ and $i=4,5,6,7$.
When the aforementioned splitting is assumed, we choose these coordinates in the {\emph{static gauge}} $x^a (\sigma ) = \sigma^a$ and such that the metric on $X$ takes the block diagonal form $g_{IJ} =( g_{ab} , g_{ij} )$. This will allow a more detailed comparison with the earlier works {\cite{harmoo, beawit}} concerning topological membrane theories. As in {\cite{harmoo, beawit}}, the four remaining coordinates are relabeled $x^i (\sigma ) = y^i (\sigma )$ and correspond to sections of the normal bundle describing fluctuations of the embedded membrane in $X$.


\subsection{\large{Choice of section}}

Our aim is to localize the path integral, defined over the  infinite-dimensional space $M = {\mbox{Map}} ( \Sigma_3 , X)$, on the finite-dimensional subspace $M_s$ of
maps corresponding to associative 3-cycles embedded in $X$.
The existence of such cycles is guaranteed by the fact that $X$ has $G_2$ holonomy.
They correspond to submanifolds ${\tilde{X}}_3 \subset X$ which are calibrated by $\Phi$ {\cite{harlaw}}.
The calibration condition follows from the relation
\be
\Phi |_{X_3} \; \leq \; {\mbox{vol}}_{X_3} \; ,
\label{assoc1}
\ee
which holds for any three-dimensional submanifold $X_3 \subset X$, with the inequality being saturated precisely for associative 3-cycles
${\tilde{X}}_3$.
Such cycles are therefore volume minimizing in their homology class.
The volume of $X_3$ in the formula above is measured with respect to the $G_2$ metric.

As noted in {\cite{harlaw, akbsal}}, an equivalent form of the associative 3-cycle equation $\Phi |_{{\tilde{X}}_3} = {\mbox{vol}}_{{\tilde{X}}_3}$ is
given by
\be
{*\Phi}_{IJKL} \, dx^J \wedge dx^K \wedge dx^L |_{{\tilde{X}}_3} \; =\; 0 \; .
\label{assoc2}
\ee
This follows from the $G_2$ identity (a list of useful $G_2$ identities is given in the appendix)
\be
{*\Phi}_{IJKL} {*\Phi}^{PQRL}  \; =\; 6\, \delta_{[I}^P \delta_J^Q \delta_{K]}^R -9\, \delta_{[I}^{[P} {*\Phi}_{JK]}^{\quad\, QR]} - \Phi_{IJK} \Phi^{PQR} \; ,
\ee
which can be used to show that
\be
{*\Phi}_{I} (u,v,w) {*\Phi}_{J} (u,v,w) g^{IJ} \; =\; || u \wedge v \wedge w ||^2 - \Phi (u,v,w)^2 \; ,
\ee
for any 3-plane defined by the three tangent vectors $u$, $v$ and $w$.
The meaning of the terms is as follows: ${*\Phi}_I (u,v,w) = {*\Phi}_{IJKL} u^J v^K w^L$,
$|| u \wedge v \wedge w ||^2 = 6\, u^{[I} v^J w^{K]} u_I v_J w_K$ and
$\Phi (u,v,w) = \Phi_{IJK} u^I v^J w^K$.
Thus since the identity above holds for all tangent spaces of a given $X_3$ then it is clear that the
right hand side (corresponding to the square of the bound ({\ref{assoc1}})) vanishes only if ({\ref{assoc2}}) is satisfied.

For the theory of maps under consideration, we can write each $X_3 = x( \Sigma_3 )$ for some map $x$ and 3-manifold $\Sigma_3$.
In terms of this embedding, it will be convenient to define a certain section of the cotangent bundle $T^* X$ via
\be
\Xi_{I} \; =\; \frac{1}{6} (*\Phi )_{IJKL} \, \partial_a x^J \partial_b x^K \partial_c x^L \, \epsilon^{abc}  \; ,
\label{assoc}
\ee
where $\epsilon^{abc}$ is the orientation tensor on $\Sigma_3$.
Clearly $\Xi_I$ vanishes only if $x( \Sigma_3 ) = {\tilde{X}}_3$ is associative and thus defines our choice of Mathai-Quillen section.
In the static gauge defined earlier, the expression ({\ref{assoc}}) can be used to write
\bea
\Xi_a &=& -\, \Phi_{ij}^{\;\;\; b} \partial_a y^i \partial_b y^j \nonumber \\
\Xi_i &=& \Phi_{ij}^{\;\;\,\, a} \partial_a y^j + \frac{1}{6} \, \Phi^{abc} {*\Phi}_{ijkl} \, \partial_a y^j \partial_b y^k \partial_{c} y^l \; .
\label{memsec}
\eea
This is derived using various identities of the associative 3-form $\Phi$ in static gauge which are summarised in the appendix
\footnote{Strictly speaking, for general $X_3 =x ( \Sigma_3 )$, $\epsilon^{abc}$ is only identified with the component $\Phi^{abc}$ up to multiplication by some nowhere vanishing function. We have factored this function into the definition of $\Xi_a$ and $\Xi_i$ above. This does not affect the structure of the theory since zeros of the redefined quantities occur at the same points in $M$ as those of the original ones (because the function has no zeros itself).}.
A refinement of the associative 3-cycle equations $\Xi_I =0$ thus follows from the fact that $\Xi_a = - \Xi_i \partial_a y^i$ and so
vanishes identically as a consequence of the four real equations $\Xi_i =0$ being satisfied (though $\Xi_a =0$ does not imply $\Xi_i =0$).
The quantity of interest is therefore $\Xi_i$ which can be understood as a section of the conormal bundle of $\Sigma_3$ in $X$ whose
zeros define associative 3-cycles in $X$.

The linearization of the map $\Xi_i$
\footnote{That is, neglecting the cubic order terms in $\Xi_i$.}
 can also be understood as a section of the bundle $E$ whose typical fibre at a point $x \in M$ is the vector space of sections
 $\Gamma ( {\sf{P}}_4 [ T^* \Sigma_3 \otimes x^* N x( \Sigma_3 )])$. The action of the projection operator ${\sf{P}}_4$ on one-forms on
 $\Sigma_3$, which are valued in the space of sections of the normal bundle of $\Sigma_3$ in $X$, is defined in the appendix.
 The isomorphism above follows from the identity
\be
\Phi_{a}^{\;\; ij} \, \Xi_j^{\mbox{\scriptsize{lin}}} \; =\; -3 \, ( {\sf{P}}_4 )_{a\;\;\, j}^{\;\; bi} \partial_b y^j   \; ,
\ee
where $\Xi_i^{\mbox{\scriptsize{lin}}} = \Phi_{ij}^{\;\;\,\, a} \partial_a y^j$.
The four linearized equations $\Xi_i^{\mbox{\scriptsize{lin}}} =0$ are therefore equivalent to the four linearly
independent \lq instanton' equations
\be
\partial_a y^i \; =\; *\Phi_{a\;\;\, j}^{\;\; bi} \partial_b y^j \; .
\label{lineq}
\ee
It should be stressed that solutions of ({\ref{lineq}}) are not guaranteed to be solutions of the full non-linear equations $\Xi_i =0$
(though $\Xi_a = - \Xi_i \partial_a y^i = - \Xi_i^{\mbox{\scriptsize{lin}}} \partial_a y^i =0$ still follow identically).
The linearized equations ({\ref{lineq}}) are easily seen to be equivalent with equations that have been studied in {\cite{mcle, akbsal}}, in the context of the deformation theory of associative 3-cycles in $G_2$ manifolds. The moduli space of associative 3-cycles generically has singularities {\cite{mcle}}.
This is to be contrasted with, for example, the moduli spaces of holomorphic curves or special Lagrangian cycles in Calabi-Yau manifolds, which are generically smooth.
The obstruction to smoothness of the moduli space of associative 3-cycles is discussed in {\cite{mcle}}
(and a certain smooth deformed moduli space proposed in {\cite{akbsal}}). The obstruction is related to the dimension of the cokernal of a so
called \lq twisted Dirac' operator (acting on harmonic \lq twisted spinors' on $\Sigma_3$)
\footnote{First order deformations $\delta y^i = \psi^i$ of ({\ref{lineq}}) lead to the equation
$\nabla_a \psi^i \; =\; *\Phi_{a\;\;\, j}^{\;\; bi} \nabla_b \psi^j$, where $\nabla_a$ is the Levi-Civita connection on $X$ pulled back to
$\Sigma_3$ via $y^i$.
Relative to our conventions and in terms of a basis of imaginary quaternions $i$, $j$, $k$ (where $i^2 = j^2 = k^2 =-1$, $ij = -ji =k$),
the twisted Dirac operator is
$\nabla = i \nabla_2 + j \nabla_1 + k \nabla_3$ whilst the harmonic twisted spinor corresponds to the quaternionic field
$\psi = \psi^6 + i \psi^4 + j \psi^5 + k \psi^7$. It is then straightforward to show that the four real equations
$\nabla_a \psi^i \; =\; *\Phi_{a\;\;\, j}^{\;\; bi} \nabla_b \psi^j$ are precisely equivalent to the single quaternionic equation $\nabla \psi =0$.}
 which is generically non-zero.

For $G_2$ manifolds that are circle fibrations over Calabi-Yau three-folds, it is straightforward to show that dimensional reduction along the
third membrane direction (which is taken to be the circular one) reduces the instanton equations above to precisely the equations defining
holomorphic curves on which the A-model localizes
\footnote{Of course, when assuming static gauge in seven dimensions, one actually obtains the 4 equations describing Riemann
surfaces whose coordinates are identified with two of the Calabi-Yau coordinates and which are holomorphically embedded in the four transverse
directions. All 6 holomorphic curve equations for the A-model follow from dimensional reduction of $\Xi_I =0$.}.
This statement is also true of the full associative 3-cycle equations $\Xi_i =0$ since the non-linear terms in ({\ref{memsec}}) vanish
identically in the reduction.


\subsection{\large{Action and BRST rules}}

Associated with our choice, $\Xi_I$, of Mathai-Quillen section are the remaining Grassmann odd variables $\psi^I$ and $\chi^I$ (up to the obvious vielbein factors), which are interpreted as fermionic fields on $\Sigma_3$.

The corresponding Mathai-Quillen action is given by
\be
I_M \; =\; \int_{\Sigma_3} d^{3} \sigma \, \left( \frac{1}{2} g^{IJ} \Xi_I \Xi_J + i \chi^I \left( \delta \Xi_I - \Gamma_{IJ}^K \psi^J \Xi_K \right) - \frac{1}{4} R_{IJKL} \psi^I \psi^J \chi^K \chi^L \right) \; ,
\label{actionm}
\ee
where
\be
\delta \Xi_I - \Gamma_{IJ}^K \psi^J \Xi_K \; =\; \frac{1}{2} \, {*\Phi}_{IJKL} \, \nabla_a \psi^J \partial_b x^K \partial_c x^L \, \epsilon^{abc}  \; ,
\ee
and $\nabla_a \psi^I = \partial_a \psi^I + \Gamma_{JK}^I \psi^J \partial_a x^K$ is obtained from the Levi-Civita connection for
$g_{IJ}$ pulled back to $\Sigma_3$ by $x$. $R_{IJKL}$ is the curvature of $g_{IJ}$.

The action ({\ref{actionm}}) is invariant under the BRST transformations
\be
\delta x^I \; =\; \psi^I \; , \quad\quad \delta \psi^I \; =\; 0  \; , \quad\quad \delta \chi^I \; =\; i g^{IJ} \Xi_J - \Gamma_{JK}^I \psi^J \chi^K \; ,
\label{brstm}
\ee
which follow from ({\ref{brst}}) and are nilpotent up to the equation of motion for $\chi^I$.
The action $I_M$ has a global $U(1)$ ghost number symmetry under which $( x , \psi , \chi )$ have charges $(0,1,-1)$.

In addition, the action $I_M$ is $\delta$-exact with
\be
I_M \; =\; \delta \Psi_M \; =\; \delta \left( -\frac{i}{2} \int_{\Sigma_3} d^{3} \sigma \, \chi^I \, \Xi_I \right) \; ,
\label{exactm}
\ee
on-shell. This property again implies that the formal Mathai-Quillen path integral should be independent of the value of a coupling constant multiplying
$I_M$ so that only the critical points of $\delta$ contribute in the large coupling limit.
From (\ref{brstm}), it is clear that these critical points precisely correspond to $\Xi_I = 0$, i.e. to membranes wrapping the associative 3-cycles of the $G_2$ manifold $X$.

It is interesting to note that $I_M$ has an additional ${\mathbb{Z}}_4$ symmetry under the transformations
\be
x^I \; \rightarrow\; x^I \; , \quad\quad \psi^I \; \rightarrow \; \chi^I  \; , \quad\quad \chi^I \; \rightarrow \; - \psi^I \; .
\ee
Performing this change of variables in the BRST rules ({\ref{brstm}}) together with the relabeling $\delta \rightarrow {\tilde{\delta}}$, one can show that the invariant $I_M$ is also closed and exact under the resulting ${\tilde{\delta}}$ transformations
\be
{\tilde{\delta}} x^I \; =\; \chi^I \; , \quad\quad {\tilde{\delta}} \chi^I \; =\; 0  \; , \quad\quad {\tilde{\delta}} \psi^I \; =\; -i g^{IJ} \Xi_J + \Gamma_{JK}^I \psi^J \chi^K \; .
\ee

One obtains a truncation of the theory above by imposing static gauge. This gauge ensures that there are no BRST variations of $x^a (\sigma) = \sigma^a$ and so $\psi^a =0$. The fermionic fields $\chi^a$ are also redundant in the quantum theory since the critical points where $\delta \chi^i =0$ imply $\delta \chi^a =0$ identically (using the identity $\Xi_a = - \Xi_i \partial_a y^i$). This observation, together with the structure above, allows us to identify the model we have found (in static gauge) with the membrane theory with two BRST charges constructed by Beasley and Witten in $(0|2)$ superspace {\cite{beawit}}. The precise expressions for their variables in terms of ours are given by $\delta_{\dot{\alpha}} = ( \delta , {\tilde{\delta}} )$, $\psi^i_{\dot{\alpha}} = ( \psi^i , 1/2 \chi^i )$, $\delta \Psi / \delta y^i = i \Xi_i$.

It is perhaps worth remarking that in static gauge one can simply replace $\Xi_i$ with its linearization $\Xi_i^{\mbox{\scriptsize{lin}}}$ in the action and BRST transformations above without losing any of the aforementioned symmetry properties of the theory.
This simpler truncated theory would localize on solutions of the instanton equations $\partial_a y^i = *\Phi_{a\;\;\, j}^{\;\; bi} \partial_b y^j$.
The BRST variation of these equations $\delta \Xi_i^{\mbox{\scriptsize{lin}}} =0$ implies
\be
\nabla_a \psi^i \; =\; *\Phi_{a\;\;\, j}^{\;\; bi} \nabla_b \psi^j \; .
\ee
The space of solutions of the equations above corresponds to the kernal of the twisted Dirac operator in {\cite{mcle}}.
As noted in {\cite{mcle}}, the dimension of this kernal is difficult to compute.
Since $\Sigma_3$ has odd dimension, the Atiyah-Singer index of the twisted Dirac operator vanishes identically.
Thus the only conclusion one can draw is that the dimensions of the kernal and cokernal of the twisted Dirac operator are equal.

One can easily check that the action and BRST transformations above dimensionally reduce to those of the A-model when one of the $\Sigma_3$ directions is identified with one of the dimensions of $X$ and taken to be the compactified circle.


\subsection{\large{Topological term}}

Similarly to the A-model case, there is an obvious topological term that one can add to the Mathai-Quillen action $I_M$.
It is given by the integral over $\Sigma_3$ of the pull back via $x$ of the associative 3-form
$\Phi = \frac{1}{6} \, \Phi_{IJK} dx^I \wedge dx^J \wedge dx^K$ on $X$, namely
\be
\int_{\Sigma_3} x^* ( \Phi ) \; =\; \int_{\Sigma_3} d^3 \sigma \, \frac{1}{6} \, \Phi_{IJK} \partial_a x^I \partial_b x^J \partial_c x^K \, \epsilon^{abc} \; .
\label{topm}
\ee
Its diffeomorphism invariance on $X$ implies that it is also BRST invariant.
Furthermore ({\ref{topm}}) depends only on the homotopy class of $x$ and the cohomology class of $\Phi$.

One can show that $\int_{\Sigma_3} x^* ( \Phi )$ exactly reduces to the term $\int_{\Sigma_2} \phi^* ( K )$ that proved so crucial
in the analysis of the A-model, under the circle compactification of the membrane mentioned above.
Indeed its inclusion would play an important role in the quantum structure of the membrane theory even though it does not
modify the classical equations of motion. Thus we argue that it should be added to the action $I_M$.

Another reason it is natural to include such a term follows by comparison with the sigma model action constructed for
physical supermembranes by Bergshoeff, Sezgin and Townsend {\cite{bersez}}.
In terms of our variables, the purely bosonic part of their action in Nambu-Goto form is given by
\be
\int_{\Sigma_3} d^3 \sigma \sqrt{{\mbox{det}} ( g_{IJ} \partial_a x^I \partial_b x^J )} \; =\; \frac{1}{2} \int_{\Sigma_3} d^3 \sigma \left( 2 + g_{ij} \partial_a y^i \partial^a y^j +... \right) \; .
\label{supmem}
\ee
The equality comes upon imposing static gauge and expanding up to quadratic order in powers of $y^i$ around an associative 3-cycle in $X$.
This expansion has been done in more detail by Harvey and Moore in {\cite{harmoo}}. The point we wish to emphasize is that, to quadratic order, such an expansion follows from the Mathai-Quillen action $I_M$ only if the topological term $\int_{\Sigma_3} x^* ( \Phi )$ (also appearing in {\cite{harmoo}}) is included. In particular, one can show that
\footnote{The term $x^* ( \Phi )$ provides the cosmological constant but also gives a term in the integrand proportional to
${*\Phi}^{ab}_{\;\;\;\, ij} \partial_a y^i \partial_b y^j$ which exactly cancels the corresponding term in
$\frac{1}{2} g^{ij} \Xi_i^{\mbox{\scriptsize{lin}}} \Xi_j^{\mbox{\scriptsize{lin}}}$.
The identity ({\ref{linid}}) can also be derived from a result used in the proof of theorem 5.3 in {\cite{mcle}}.}
\be
\int_{\Sigma_3} d^3 \sigma \left( \frac{1}{2} g^{ij} \Xi_i^{\mbox{\scriptsize{lin}}} \Xi_j^{\mbox{\scriptsize{lin}}} + x^* ( \Phi ) \right) \; =\; \frac{1}{2} \int_{\Sigma_3} d^3 \sigma \left( 2+ g_{ij} \partial_a y^i \partial^a y^j \right) \; .
\label{linid}
\ee
%


\subsection{\large{Observables}}

Formally at least, we can proceed in a similar manner as for the A-model and consider the observables
\be
\langle {\cal{O}}_{p_1} ... {\cal{O}}_{p_s} \rangle (t) \; =\; \int_M [d x ] \, [d \psi ] \, [d \chi ] \, {\cal{O}}_{p_1} ... {\cal{O}}_{p_s} \, e^{-t S_M}
\label{obs2}
\ee
of the BRST invariant local operators $\{ {\cal{O}}_{p_r} | r=1,...,s \}$ with ghost numbers $p_r$ , which are evaluated at points $\sigma_r \in \Sigma_3$ such that $x ( \sigma_r )$ lies on a homology $p_r$-cycle in $X$. The path integral above is with respect to the full action
\be
t\, S_M \; =\; t\, I_M + t \int_{\Sigma_3} x^* ( \Phi ) \; ,
\ee
where $t$ is the coupling constant.

For a $G_2$ manifold with a fixed associative 3-form $\Phi$, one therefore expects the path integral ({\ref{obs2}}) to reduce to a sum over homotopy classes $[x]$ of the maps $x$. Each term in the sum, corresponding to a Mathai-Quillen type path integral
\be
\langle {\cal{O}}_{p_1} ... {\cal{O}}_{p_s} \rangle_{[x]} \; =\; \int_{M_{[x]}} [dx] \, [d \psi ] \, [d \chi ] \, {\cal{O}}_{p_1} ... {\cal{O}}_{p_s} \, e^{-t I_M} \; ,
\label{resobs2}
\ee
is restricted to only that portion of configuration space $M_{[x]} \subset M$ which corresponds to maps $x$ in a given homotopy class $[x]$ (and obeying $x ( \sigma_r ) \in H_{p_r} (X,{\mathbb{R}})$). Formal arguments imply that ({\ref{resobs2}}) is independent of $t$ since it involves only the BRST-exact part $I_M$ of the full action $S_M$. The topological term is factored out of the restricted path integral above since it takes the same value for all maps in the same homotopy class. Thus it weighs each term ({\ref{resobs2}}) by a {\emph{membrane instanton}} factor
\be
e^{-t \int_{\Sigma_3} x^* ( \Phi )}
\ee
in ({\ref{obs2}}).
The coupling constant independence of ({\ref{resobs2}}) has been used to localize this path integral on associative 3-cycles, such that $x ( \Sigma_3 ) = {\tilde{X}}_3 \subset X$ in the subspace $M_{[x]}$. Thus the above factor can be written as $e^{-t\, \Phi |_{{\tilde{X}}_3}}$ , which is in precise agreement with the contribution from isolated membrane instantons proposed in \cite{harmoo}.


\subsubsection{Local operators and $G_2$ cohomology}

The classification of observables in this theory simply involves computing the cohomology of $\delta$.
This can be achieved via a straightforward generalization of the techniques used in the analysis of the A-model.
In particular, one has the simple formula relating the actions of $\delta$ on local operators and $d$ on forms on $X$
\be
\delta {\cal{O}}_{\omega_p} \; =\; {\cal{O}}_{d \omega_p } \; ,
\ee
where the local operator ${\cal{O}}_{\omega_p} = 1/p! \, \omega_{I_1 ... I_p} (x) \psi^{I_1} ... \psi^{I_p}$ at the point $\sigma \in \Sigma_3$,
$x (\sigma ) \in X$ and $\omega_p = 1/p! \, \omega_{I_1 ... I_p} (x) dx^{I_1} \wedge ... \wedge dx^{I_p}$.
One can construct a basis of BRST invariant local operators with ghost number $p$ from the set of closed $p$-forms one
obtains as Poincar\'{e} duals of the corresponding homology $p$-cycles in $X$
(delta function support of such operators on the dual homology cycles is thus required). This correspondence is an isomorphism.

The de Rham cohomology groups on the $G_2$ manifold $X$ have the following decompositions
\bea
H^0 (X,{\mathbb{R}}) &=& {\mathbb{R}} \nonumber \\
H^1 (X,{\mathbb{R}}) &=& H^1_{\bf{7}} (X,{\mathbb{R}}) \nonumber \\
H^2 (X,{\mathbb{R}}) &=& H^2_{\bf{7}} (X,{\mathbb{R}}) \oplus H^2_{\bf{14}} (X,{\mathbb{R}}) \nonumber \\
H^3 (X,{\mathbb{R}}) &=& H^3_{\bf{1}} (X,{\mathbb{R}}) \oplus H^3_{\bf{7}} (X,{\mathbb{R}}) \oplus H^3_{\bf{27}} (X,{\mathbb{R}}) \; .
\eea
Similar decompositions follow for the remaining cohomology groups by Hodge duality and will therefore be ignored in our analysis. The subscripts in $H^I_{\bf{n}}$ denote the irreducible representations ${\bf{n}}$ of $G_2$ that the $I$-form components occupy. The non-trivial projection operators ${\sf{P}}^I_{\bf{n}}$ onto these irreducible subspaces are given by
\bea
( {\sf{P}}^2_{\bf{7}} )_{IJ}^{\;\;\;\, PQ} &=& \frac{1}{6} \, \Phi_{IJA} \Phi^{PQA} \; =\; \frac{1}{3} \, \left( \delta_{[I}^P \delta_{J]}^{Q} + \frac{1}{2} \, {*\Phi}_{IJ}^{\;\;\;\, PQ} \right) \nonumber \\
( {\sf{P}}^2_{\bf{14}} )_{IJ}^{\;\;\;\, PQ} &=& \delta_{[I}^P \delta_{J]}^{Q} -\frac{1}{6} \, \Phi_{IJA} \Phi^{PQA} \; =\; \frac{2}{3} \, \left( \delta_{[I}^P \delta_{J]}^{Q} - \frac{1}{4} \, {*\Phi}_{IJ}^{\;\;\;\, PQ} \right) \nonumber \\
( {\sf{P}}^3_{\bf{1}} )_{IJK}^{\quad\;\, PQR} &=& \frac{1}{42} \, \Phi_{IJK} \Phi^{PQR} \nonumber \\
( {\sf{P}}^3_{\bf{7}} )_{IJK}^{\quad\;\, PQR} &=& \frac{1}{24} \, {*\Phi}_{IJKA} {*\Phi}^{PQRA} \nonumber \\
( {\sf{P}}^3_{\bf{27}} )_{IJK}^{\quad\;\, PQR} &=& \delta_{[I}^P \delta_J^Q \delta_{K]}^R - \frac{1}{42} \, \Phi_{IJK} \Phi^{PQR} -\frac{1}{24} \, {*\Phi}_{IJKA} {*\Phi}^{PQRA} .
\eea
These can be checked using the $G_2$ identities in the appendix.

Smooth compact $G_2$ manifolds have a somewhat simpler cohomology due to the fact that all $H^I_{\bf{7}} =0$ (that is when the holonomy is the full $G_2$ and not a proper subgroup thereof). The only independent non-trivial cohomology groups in this case are $H^3_{\bf{1}}$, $H^3_{\bf{27}}$ and $H^2_{\bf{14}}$. A useful way to analyze the first two is to observe the isomorphism
\be
\alpha_{IJK} \; =\; 3 \Phi_{[IJ}^{\;\;\;\;\, A} \xi_{K]A} \; ,
\ee
between the components $\alpha_{IJK}$ in $\Lambda^3_{\bf{1}} \oplus \Lambda^3_{\bf{27}}$ and the symmetric tensor representation $\xi_{IJ} = \xi_{JI}$ of $G_2$. The traceless part $\xi_{IJ} - \frac{1}{7} \, g_{IJ} \xi^K_{\; K}$ of $\xi_{IJ}$ is isomorphic to $\Lambda^3_{\bf{27}}$ whilst its trace part $\frac{1}{7} g_{IJ} \xi^K_{\; K}$ is isomorphic to the singlet representation $\Lambda^3_{\bf{1}}$. Thus the only elements of $H^3_{\bf{1}}$ are constant multiples of $\Phi$. Furthermore one can show that if the 3-form $\alpha$ defined above is closed and coclosed (i.e. harmonic) then it follows that $\xi$ obeys
\be
\xi^I_{\; I} \; =\; 0 \; , \quad \nabla^I \xi_{IJ} \; =\; 0 \; , \quad \Phi_{I}^{\;\; KL} \nabla_K \xi_{LJ} \; =\; 0 \; .
\ee
A nice observation in {\cite{DWW}} is that the equations above are precisely those satisfied by the small variation $\xi_{IJ} = \delta g_{IJ}$ of a $G_2$ holonomy metric $g_{IJ}$ in order that the new metric $g_{IJ} + \delta g_{IJ}$ also has $G_2$ holonomy. These equations have also been used recently in the analysis of physical states for the topological $G_2$ string in {\cite{BNS}}. Thus elements of $H^3_{\bf{27}}$ correspond to such $G_2$ holonomy preserving deformations. Finally, any element of $\Lambda^2_{\bf{14}}$ can be written as ${\sf{P}}^2_{\bf{14}} \beta$ for some 2-form $\beta$ on $X$. Such elements have no other special properties, to the best of our knowledge, except that closure $d\, ({\sf{P}}^2_{\bf{14}} \beta) =0$ of ${\sf{P}}^2_{\bf{14}} \beta$ implies coclosure $d^\dagger ({\sf{P}}^2_{\bf{14}} \beta) =0$ identically.

In conclusion, for smooth compact manifolds with $G_2$ holonomy, the de Rham cohomology is spanned by the three kinds of linearly independent harmonic forms $\Phi = {\sf{P}}^3_{\bf{1}} \Phi$, $\alpha = {\sf{P}}^3_{\bf{27}} \alpha$ and $\beta = {\sf{P}}^2_{\bf{14}} \beta$ mentioned above. The corresponding BRST invariant local operators are
\bea
{\cal{O}}_{\Phi} &=& \frac{1}{6} \, \Phi_{IJK} \psi^I \psi^J \psi^K \nonumber \\
{\cal{O}}_{\alpha} &=& \frac{1}{2} \, \Phi_{IJ}^{\;\;\;\;\, A} \xi_{KA} \psi^I \psi^J \psi^K \nonumber \\
{\cal{O}}_{\beta} &=& \frac{1}{2} \, \beta_{IJ} \psi^I \psi^J \; .
\eea
Of course, for more general manifolds with holonomy in a subgroup of $G_2$ one can have additional local operators corresponding to elements of $H^I_{\bf{7}}$.


\subsubsection{Non-local operators}

Non-local BRST invariant operators ${\cal{O}}^{(n)} = \int_{\gamma_n} W_n$ defined on homology $n$-cycles
$\gamma_n$ in $\Sigma_3$ can be constructed from a local BRST invariant operator ${\cal{O}} = W_0$ via the descent equations
\be
\delta W_n \; =\; d W_{n-1} \; ,
\ee
which define the set of $n$-forms $W_n$ on $\Sigma_3$ for $n=1,2,3$.

The local BRST invariant operator ${\cal{O}}_\Phi = \frac{1}{6} \, \Phi_{IJK} \psi^I \psi^J \psi^K$ associated with $\Phi$ has the following non-local descendents
\bea
{\cal{O}}^{(1)}_\Phi &=& \int_{\gamma_1} \frac{1}{2} \, \Phi_{IJK} \, \partial_a x^I \psi^J \psi^K \, d\sigma^a \nonumber \\
{\cal{O}}^{(2)}_\Phi &=& \int_{\gamma_2} - \frac{1}{2} \, \Phi_{IJK} \, \partial_a x^I \partial_b x^J \psi^K \, d \sigma^a \wedge d \sigma^b \\
{\cal{O}}^{(3)}_\Phi &=& \int_{\Sigma_3} - \frac{1}{6} \, \Phi_{IJK} \, \partial_a x^I \partial_b x^J \partial_c x^K \, d \sigma^a \wedge d \sigma^b \wedge d \sigma^c \; =\; - \int_{\Sigma_3} x^* ( \Phi )  \; . \nonumber
\eea
A similar set of descendents
\bea
{\cal{O}}^{(1)}_{\alpha} &=& \int_{\gamma_1} \frac{1}{2} \, \left( 2\, \Phi_{IJ}^{\;\;\;\;\, A} \xi_{KA} + \Phi_{JK}^{\;\;\;\;\, A} \xi_{IA} \right) \, \partial_a x^I \psi^J \psi^K \, d\sigma^a \nonumber \\
{\cal{O}}^{(2)}_{\alpha} &=& \int_{\gamma_2} - \frac{1}{2} \, \left( \Phi_{IJ}^{\;\;\;\;\, A} \xi_{KA} + 2\, \Phi_{KI}^{\;\;\;\;\, A} \xi_{JA} \right) \, \partial_a x^I \partial_b x^J \psi^K \, d \sigma^a \wedge d \sigma^b \\
{\cal{O}}^{(3)}_{\alpha} &=& \int_{\Sigma_3} - \frac{1}{2} \, \Phi_{IJ}^{\;\;\;\;\, A} \xi_{KA} \, \partial_a x^I \partial_b x^J \partial_c x^K \, d \sigma^a \wedge d \sigma^b \wedge d \sigma^c \; =\; - \int_{\Sigma_3} x^* ( \alpha )  \; , \nonumber
\eea
follows from the local BRST invariant operators ${\cal{O}}_{\alpha}$. Finally, the local BRST invariant operators ${\cal{O}}_\beta$ lead to the descendents
\bea
{\cal{O}}^{(1)}_{\beta} &=& \int_{\gamma_1} \beta_{IJ} \, \partial_a x^I \psi^J \, d\sigma^a \nonumber \\
{\cal{O}}^{(2)}_{\beta} &=& \int_{\gamma_2} - \frac{1}{2} \, \beta_{IJ} \, \partial_a x^I \partial_b x^J \, d \sigma^a \wedge d \sigma^b \; =\; - \int_{\gamma_2} x^* ( \beta )  \; .
\eea
Notice that ${\cal{O}}^{(1)}$ need not be trivial even for compact $G_2$ holonomy manifolds since $H^1 (X,{\mathbb{R}}) =0$ does not generically imply that $H^1 ( \Sigma_3 ,{\mathbb{R}})$ is trivial (the former statement is only a restriction on the allowed embeddings of homology cycles on $\Sigma_3$ in $X$).


\section{\large{Concluding remarks}} \label{sec-cr}
\setcounter{equation}{0}

In the present paper we constructed a topological theory of membranes wrapping associative three-cycles in a $G_2$ manifold, using the Mathai-Quillen approach.
This description of the theory and its observables should be considered as first steps towards formulating a full topological membrane theory. There are evidently many open directions and various issues which require further investigation. We conclude by discussing some of them in more detail.


\subsection{Membrane instanton expansion}

The topological membrane theory we have described can be viewed as a generalization of the topological A-model,
where the string worldsheet is replaced by the membrane worldvolume, and localization on
holomorphic curves in ${\mbox{CY}}_3$ is substituted with localization on associative 3-cycles in the $G_2$ manifold.
Recall that the free energy $F$ of the A-model of topological strings can be expressed as a sum over worldsheets of genus $g$ such that $F = \sum_{g=0}^\infty \lambda^{2-2g} F_g$ where $\lambda$ is the string coupling and the coefficient $F_g$ is the genus $g$ free energy. The genus zero free energy has the following expansion
\be
F_0 \; =\; \int_{X} K \w K \w K + \sum_{\Sigma \in H_2(X, \mathbb{Z})} d_\Sigma \sum_{m=1}^{\infty} \frac{1}{m^3} e^{- m\, K|_{\Sigma}} \; ,
\ee
where $X = {\mbox{CY}}_3$ and $\Sigma$ is the homology class of a holomorphic curve in $X$ \cite{nv}.
The first term in $F_0$ is the volume of the Calabi-Yau corresponding to the contribution to the free energy from constant maps into $X$
\footnote{Constant maps take the whole worldsheet to a single point in $X$. Thus summing over such maps is effectively counting the points in $X$, i.e. computing the volume.}.
The remaining terms correspond to worldsheet instantons which are summed over all homology classes with degeneracy $d_\Sigma$ and weighed by the
exponentiated area of the holomorphic curve in $X$. The summation over $m$ represents multiply wrapped worldsheets with appropriate degeneracy factor $1/ m^3$.

It is natural to ask whether, in principle, the topological membrane theory could have a similar notion of \lq genus expansion'.
Of course, for closed oriented string worldsheets, the situation is much clearer since the genus classifies their topology.
The classification of closed oriented 3-manifold topologies for membrane worldvolumes is much more difficult though.
There is indeed a notion of genus for such 3-manifolds associated with their Heegaard splitting
\footnote{Every closed orientable 3-manifold $\Sigma_3$ can be decomposed topologically in terms of two handlebodies
$H_g \cup H^\prime_g$ (for some genus $g$) whose boundaries coincide $\partial H_g = \partial H^\prime_g$.
A handlebody $H_g$ of genus $g$ is a closed orientable Riemann surface of genus $g$ with the interior points filled in
(the Riemann surface is therefore $\partial H_g$). However, this decomposition is not unique: the existence of a Heegaard
splitting $\Sigma_3 = H_g \cup H^\prime_g$ implies that there is also a splitting $\Sigma_3 = H_{g+1} \cup H^\prime_{g+1}$.
The minimum $g$ for which the decomposition exists is called the {\emph{Heegaard genus}} of $\Sigma_3$.}.
However, there are many different 3-manifold topologies with the same Heegaard genus and therefore the latter does not
provide a classification (only for Heegaard genus zero is there a unique 3-manifold $S^3$).

In fact the classification of 3-manifold topologies is intimately related to the classification of knots and links in $S^3$.
This relationship follows from a beautiful theorem due to Lickorish and Wallace {\cite{LW}} which states that every closed oriented 3-manifold topology
can be obtained via surgery on some framed link in $S^3$
\footnote{The framing of a link corresponds to an integer assigned to each knot component of the link.
Each such integer can be understood as the number of times a slightly displaced copy of the associated knot winds around the original knot.
A knot component and its frame are contained in a region of $S^3$ homeomorphic to a solid torus.
The surgery begins by removing all such solid tori associated with a given framed link in $S^3$.
The framing of each knot component then determines the number of Dehn twists one performs on each torus before
gluing them back in the mutilated $S^3$, to obtain a new 3-manifold topology.
The theorem states that all closed oriented 3-manifold topologies can be obtained in this way.}.
This is not to say that the correspondence is one-to-one and indeed there are many different framed links that give rise to the same 3-manifold topology.
One can however make use of a theorem due to Kirby {\cite{Kir}}, stating that the 3-manifolds obtained by surgery on two different framed links are
homeomorphic only if the two links are related by a sequence of Kirby moves
\footnote{There are two types of Kirby moves.
The first type adds or removes a disjoint unknotted circle with framing $\pm 1$ to or from a given link.
The second type joins together two knot components of a given link by taking the connected sum of (i.e. attaching a band connecting)
the frame of the second knot with the first knot itself.
These moves change the isotopy class of a given knot or link but not the topology of the resulting 3-manifold.}.
Of course, this structure only allows one to map the classification of 3-manifold topologies to the classification of knots and links
that are identified under Kirby moves, which is itself an unsolved problem.
Indeed even determining whether two arbitrarily complicated links are related by Kirby moves can be very difficult.
Nonetheless it seems an intriguing prospect to arrange an expansion in closed membrane worldvolume topologies as a sum over knots and links
modulo Kirby moves.

Another, more direct approach to the difficult problem of classifying compact oriented 3-manifolds is provided by the Thurston geometrization conjecture \cite{WT} thought to have been proven only recently by Perelman \cite{GP}
\footnote{As far as we are informed this conjecture is believed to be correct, however Perelman's proof is still under investigation.
Also, the classification of higher-dimensional manifolds in full extent is known to be impossible, but there is a classification for
simply connected manifolds.}.
Classical decomposition theorems state that any compact oriented 3-manifold can be expressed as the connected sum of a unique
(up to homeomorphism) collection of prime 3-manifolds which are topologically either $S^2 \times S^1$ or irreducible (meaning any embedded
$S^2$ bounds a 3-ball in the prime 3-manifold).
Irreducible 3-manifolds have a further canonical factorization into components which are divided by (disjointly embedded) incompressible tori.
The Thurston conjecture states that the interior of each component 3-manifold in the decompositions above has the local geometry of a
homogeneous space with finite volume. Furthermore, the latter space is locally isometric  to a finite volume quotient by a freely acting discrete isometry group of one of eight spaces, corresponding to the flat ${\mathbb{R}}^3$,
hyperbolic $H^3$, elliptic $S^3$, $H^2 \times {\mathbb{R}}$, $S^2 \times {\mathbb{R}}$, $SL(2,{\mathbb{R}})$, Nilpotent and Solve geometries
\footnote{Nilpotent geometry is another name for the Heisenberg group manifold whose elements are upper triangular $3\times 3$ matrices with
unit entries along the diagonal.
The Lie algebra of this group is just the Heisenberg algebra that appears in quantum mechanics and is therefore nilpotent because repeated
commutators of general elements in the algebra vanish.
Solve geometry is the group manifold ${\mathbb{R}} \ltimes {\mathbb{R}}^2$ whose elements are upper triangular $2\times 2$ matrices.}.
It is therefore conceivable that the free energy of the membrane theory could be written as a sum over different 3-manifold topologies,
arranged according to the factorization properties above, with each term being integrated over the moduli space of
possible local geometries appearing on Thurston's list.

As in the case of the A-model, the simplest contribution to the free energy of the membrane theory would come from constant maps.
These correspond to trivial (i.e. contractible) 3-cycles mapping the whole worldvolume of the membrane to a point in the $G_2$ manifold.
We would again naturally expect such contributions to \lq count points' on the $G_2$ manifold $X$, giving its volume.
That is, one anticipates the contribution to be given by the \lq on-shell' Hitchin action
\be
F_{\mbox{\scriptsize{const}}}^{G_2} \; =\; \int_{X}  \Phi \w * \Phi \; .
\ee
Of course, the Hitchin action above is to be expected from the classical (large $G_2$ volume) approximation of topological
M-theory \cite{M} but also, via the membrane-string duality in seven dimensions, from its appearance as the genus zero free
energy of the topological $G_2$ string \cite{BNS}.
In addition, it is natural to expect that the additional contributions to the free energy from membrane instantons would be weighed by the factor
\be
e^{- \int_{\Sigma_3} x^* (\Phi)}
\ee
which would give back the topological A-model worldsheet instanton expansions upon dimensional reduction of the membrane on a circle.

It is amusing to speculate that the degeneracy of membranes wrapping $m$ times over an associative 3-cycle
(i.e. weighing the proposed instanton sum by $e^{-m \int_{\Sigma_3} x^* (\Phi)}$) is proportional to $m^{-7/3}$.
Here the power of $m$ is inversely related to the degree of the volume of the $G_2$ manifold, understood as a homogeneous polynomial in 3-form $\Phi$
of degree $7/3$.
Such a conjecture is consistent with the single power of both $m$ and $\Phi$ in the instanton factor exponent.
The proposal is motivated via naive analogy with the A-model, where the degeneracy $m^{-3}$ of A-model worldsheets wrapped $m$ times over
holomorphic curves can be inferred from the inverse scaling of the Calab-Yau volume $\int K \w K\w K$ in the 2-form $K$.
Without performing the appropriate membrane path integrals explicitly, however, the proposals above remain merely conjectures. A good starting point for studying the membrane instanton expansion is
restricting to 3-manifolds which are circle fibration of a Riemann surface. In this case the 3-manifold
topologies can be classified easily and that may provide a handle on finding the degeneracy
factors in the membrane instanton expansion.

We leave a detailed analysis of the membrane instanton expansion for future work.
A possible way to approach this problem is to make use of the target space Gopakumar-Vafa point
of view of topological amplitudes \cite{gv}. The relation of our proposed membrane
instanton expansion to the Gopakumar-Vafa expansion may be an M-theory lift of the
relation between the A-model worldsheet theory and the target space
Donaldson-Thomas description.


\subsection{Additional open questions}

Perhaps the most immediate open problem is to work out the relation of the
topological membrane formulation to that of the topological $G_2$ string \cite{BNS}.
By Hodge duality in seven dimensions we expect the two formalisms to be dual to each other\footnote{That is, a 3-form gauge potential which couples to the membrane worldvolume has a 4-form field strength which
is Hodge-dual to the 3-form field strength of a 2-form potential that couples to a string worldsheet.}.
At first sight, the relation between them is similar to 
the one between the topological A- and B-models. The topological membrane theory
is the analogue of the A-model (and consistent with that, it is expected  to have a membrane
instanton expansion, similarly to the Gromov-Witten theory of the A-model).
On the other hand the topological $G_2$ string localizes on constant maps, and from
this point of view it is like the B-model, where the structure of
the theory is encoded in the special geometry.

However, there is an important difference which we should point out between the
topological membrane and $G_2$ string formulations.
Compactifying the target space of the topological $G_2$ string on
${\mbox{CY}}_3 \times S^1$ reduces the theory to a combination of the  topological A- and B-models\footnote{To be precise, this is really a combination of $A + {\bar A} + B + {\bar B}$ models.
The bar denotes a ${\mathbb{Z}}_2$ conjugation of the theories which is trivial classically. But it can be important
to distinguish between the conjugate versions at the quantum level. This distinction will not affect our present analysis.}.
We conjectured that the leading term in the topological membrane effective action is the Hitchin functional,
which is identical to the leading term in the $G_2$ string effective action.
In the classical (large $G_2$ volume) limit, both theories should therefore reduce to the same combination of A- and B-models. Indeed, the reduction performed in \cite{BNS} is purely based on the classical $G_2$ geometry.
From the worldvolume point of view however there are two sectors of the compactified membrane theory.
We found that membranes wrapping the compactified circle reduce precisely to the perturbative content of the A-model due to the reduction of the 
associative 3-cycles wrapped by them to holomorphic curves, which are  wrapped by topological strings.
On the other hand, membranes not wrapping the compactified circle reduce to 3-dimensional branes in the CY$_3$ geometry, which have natural
interpretation as Lagrangian branes in the non-perturbative sector of the A-model or maybe as NS2-branes of B-model.
We postpone a more detailed analysis of these branes for future work.

It would be important to clarify the relationship of the B-model
and the membrane formalism. In this regard, we note the interesting fact 
that Rozansky-Witten theory \cite{RW} with 4-dimensional hyperK\"{a}hler target
space can
be understood as a static gauge membrane theory of maps from 3-manifolds to
the target space. In particular, dimensional reduction along a membrane
direction gives rise to the B-model, as was first observed in \cite{thomps}.

The proposed electric-magnetic duality relation with the topological $G_2$ string could
also help to determine the measure in the topological membrane
theory in the formal expressions for our correlators. It may be
feasible to extract information about the measure on the membrane side from
the topological $G_2$ string.
A related open issue is coupling the topological membrane theory to topological gravity.

Another significant question which we have not addressed yet is the inclusion
of a background 3-form field. The latter is inherited from the $C$-field of eleven-dimensional supergravity and couples to the membrane world-volume.
The effect of the $C$-field on the topological term can be easily inferred from
the Bergshoeff-Sezgin-Townsend supermembrane action to be $\int_{\Sigma_3} x^*(\Phi + iC)$ (this is also noted in \cite{harmoo}).
Since $\int_{\Sigma_3} x^*(\Phi )$ reduces exactly to $\int_{\Sigma_2} x^*(K)$ on ${\mbox{CY}}_3 \times S^1$
(where $S^1$ is identified with a membrane direction), then it is clear that $\int_{\Sigma_3} x^*(\Phi + iC)$ reduces to
$\int_{\Sigma_2} x^*(K+iB)$ for 2-form $B$-field which complexifies the K\"{a}hler form $K$.

However, in a non-zero background $C$-field, the target space of the membrane sigma model would have to be a manifold
with $G_2$ structure instead of $G_2$ holonomy.
In such a case it is not obvious how to choose an appropriate section in the Mathai-Quillen approach.
In this regard, it may be helpful to use some of the various notions of generalized calibrations like those introduced in \cite{GIP} and \cite{PK}.
We leave that for future research.

Let us also note that one could attempt to obtain perturbative information about topological M-theory at the one-loop level by
applying an approach like the one of \cite{PW} to the second variation of the $G_2$ Hitchin functional, that has been computed in section
7.2 of \cite{NH}.

Finally, it is also intriguing to ask the question whether there is a formulation (possibly in some limit) of topological M-theory in
terms of D0-branes similarly to the Matrix description of physical M-theory.

We hope to come back to these interesting problems in the near future.


\section*{\large{Acknowledgments}}

We would like to thank David Berman, Jan de Boer, Andreas Brandhuber, Sergei
Gukov, Asad Naqvi and Bill Spence for helpful discussions related to this
work. The work of LA and PdM is supported in part by DOE grant
DE-FG02-95ER40899. The work of AS is partially supported by Stichting FOM.


\appendix

\section{\large{$G_2$ identities}}
\setcounter{equation}{0}

A seven-dimensional Riemann manifold $X$ is guaranteed to have holonomy in the subgroup $G_2 \subset SO(7)$ by the existence of a harmonic 3-form $\Phi$.
In our conventions $\Phi$ and its Hodge-dual $*\Phi$ can be written
\bea
\Phi &=& e^{123} - e^{147} - e^{156} - e^{246} + e^{257} + e^{345} + e^{367} \nonumber \\
*\Phi &=& e^{1245} + e^{1267} + e^{1346} - e^{1357} - e^{2347} - e^{2356} + e^{4567} \label{orth} \; ,
\eea
with respect to an orthonormal basis $e^I$ (where $e^{I_1 ... I_p} = 1/p! \,  e^{I_1} \wedge ... \wedge e^{I_p}$).

Some useful identities for products of the components of $\Phi$ and $*\Phi$ are as follows
\bea
{*\Phi}_{IJKA} {*\Phi}^{PQRA} &=& 6\, \delta_{[I}^P \delta_J^Q \delta_{K]}^R -9\, \delta_{[I}^{[P} {*\Phi}_{JK]}^{\;\;\;\;\; QR]} - \Phi_{IJK} \Phi^{PQR} \nonumber \\
{*\Phi}_{IJKA} \Phi^{PQA} &=& -6\, \delta_{[I}^{[P} \Phi_{JK]}^{\;\;\;\;\;\, Q]} \nonumber \\
\Phi_{IJA} \Phi^{PQA} &=& 2\, \delta_{[I}^P \delta_{J]}^Q + {*\Phi}_{IJ}^{\;\;\;\; PQ} \; .
\eea
All other required identities follow by taking contractions of the ones above. These identities can be proven in the orthonormal basis above but it is clear they are also valid in any coordinate basis by simply acting on the formulae with the appropriate combination of vielbeins.

Notice that the static gauge components $\Phi_{abi}$, $\Phi_{ijk}$, $*\Phi_{abci}$ and $*\Phi_{aijk}$ are all identically zero in ({\ref{orth}}) (and indeed in any other coordinate system that respects the splitting of $TX$ into tangent and normal bundle of $x (\Sigma_3 )$ in $X$).
The components $\Phi_{abc}$, $*\Phi_{ijkl}$ have one non-vanishing unit coefficient each and define the orientations of the
tangent and normal bundle of $x (\Sigma_3 )$ in $X$ in static gauge.
The remaining components $\Phi_{aij}$ and $*\Phi_{abij}$ each have six independent non-vanishing unit coefficients.
Note that $\Phi_{abc}$, $\Phi_{aij}$, $*\Phi_{ijkl}$ and $*\Phi_{abij}$ can be more general functions on $X$ when transformed to other coordinate
systems respecting the above decomposition of $TX$.

Some further useful identities for products of $\Phi$ and $*\Phi$ components in static gauge are
\bea
\Phi_{aij} \Phi^{bcd} &=& \delta_a^b {*\Phi}_{\;\;\,\, ij}^{cd} + \delta_a^c {*\Phi}_{\;\;\,\, ij}^{db} + \delta_a^d {*\Phi}_{\;\;\,\, ij}^{bc}  \nonumber \\
\Phi_{aik} \Phi^{bjk} &=& \delta_a^b \delta_i^j - *\Phi_{a \; i}^{\;\; b\; j} \nonumber \\
\Phi_{aml} \, {*\Phi}^{ijkl} &=& - \delta_m^i \Phi^{jk}_{\;\;\,\, a} - \delta_m^j \Phi^{ki}_{\;\;\,\, a} - \delta_m^k \Phi^{ij}_{\;\;\,\, a}  \nonumber \\
*\Phi_{acik} \, {*\Phi}^{bcjk} &=& 2 \, \delta_a^b \delta_i^j - {*\Phi}_{a \; i}^{\;\; b\; j}  \; ,
\label{iden}
\eea
which again hold in any coordinate system respecting the decomposition of $TX$ into tangent and normal bundles.

The components $*\Phi_{abij}$ can be understood as linear maps acting on the 12-dimensional vector space spanned by elements of the form $v_a^i$ (written $v \rightarrow *\Phi v$). Technically speaking this vector space is isomorphic to the tensor product space of one-forms on $\Sigma_3$ with sections of the normal bundle of $\Sigma_3$ in $X$. The last equation in ({\ref{iden}}) can then be interpreted as the quadratic equation $(*\Phi )^2 + *\Phi -2 = 0$ for the operator $*\Phi_{abij}$ acting on this space. The two roots 1 and -2 of this equation allow one to construct projection operators onto two irreducible subspaces of the aforementioned tensor product space. These projectors are
\be
{\sf{P}}_4 \; =\; \frac{1}{3} \left( 1 - *\Phi \right) \; , \quad\quad {\sf{P}}_8 \; =\; \frac{1}{3} \left( 2 + *\Phi \right) \; ,
\ee
which obey $( {\sf{P}}_4 )^2 = {\sf{P}}_4$, $( {\sf{P}}_8 )^2 = {\sf{P}}_8$ and ${\sf{P}}_4 {\sf{P}}_8 = {\sf{P}}_8 {\sf{P}}_4 = 0$ as required. The subscripts 4 and 8 denote the dimensions of the orthogonal subspaces of the 12-dimensional space onto which they project.


\end{document}